\documentclass[%
 reprint,
 amsmath,amssymb,
 aps,
pra,
]{revtex4-2}

\usepackage{graphicx}
\usepackage{dcolumn}
\usepackage{bm}
\usepackage{xcolor}
\usepackage[hidelinks]{hyperref}
\hypersetup{
    colorlinks,
    linkcolor={red!50!black},
    citecolor={blue!50!black},
    urlcolor={blue!80!black}
}

\begin{document}

\preprint{APS/123-QED}

\title{Equivalent Electric Model of a Macrospin}

\author{Steven Louis}
\email{slouis@oakland.edu}
\author{Hannah Bradley}
\email{hbradley@oakland.edu}
\author{Vasyl Tyberkevych}%
 \email{tyberkev@oakland.edu}

\affiliation{Oakland University}%

\date{\today}

\begin{abstract}
Dynamics of a ferromagnetic macrospin (e.g., a free layer of a magnetic tunnel junction (MTJ)) can be described in terms of equivalent capacitor charge $Q$ and inductor flux $\Phi$, in a manner similar to a standard electric LC circuit, but with strongly nonlinear and coupled capacitance and inductance.
This description allows for the inclusion of Gilbert damping and spin transfer torques and yields a relatively simple equivalent electric circuit, which can be easily modeled in LTspice or other electrical engineering software.
It allows one to easily simulate advanced electrical circuits containing MTJs and conventional electronic components in standard simulation software.

\end{abstract}

\maketitle

\section{Introduction}

The study of magnetization dynamics in nanoscale ferromagnetic structures is essential for advancing spintronic technologies, including magnetic tunnel junctions (MTJs) and spin-transfer torque (STT)-based devices.  
A key challenge in analyzing these systems lies in their inherently nonlinear and coupled nature, governed by the Landau-Lifshitz-Gilbert-Slonczewski (LLGS) equation.  
Traditional approaches to modeling magnetization dynamics often rely on solving this equation numerically, which can be computationally expensive and difficult to integrate into broader circuit-based simulations.  

To address this challenge, we propose an alternative approach that reformulates macrospin dynamics using an equivalent electrical circuit model.  
By establishing a mapping between the macrospin motion and the charge-flux description of an LC circuit, we derive an intuitive representation that allows for a direct correspondence between magnetization dynamics and standard circuit elements such as capacitance, inductance, and resistance.  
This formalism not only simplifies the analysis of spintronic systems, but also enables their simulation using conventional electrical engineering tools, such as LTspice.  

In this work, we first review the charge-flux representation for a standard LC circuit, providing a foundation for applying the same formalism to a ferromagnetic macrospin.  
We then extend this framework to derive an equivalent circuit for the macrospin, incorporating effects such as Gilbert damping and spin-transfer torque through resistive and source elements.  
Finally, we apply this approach to the case of an easy-plane MTJ, demonstrating how the energy landscape and spin-transfer interactions can be effectively captured using circuit components.  
By bridging the gap between magnetization dynamics and circuit theory, this work provides a powerful and practical methodology for analyzing and designing spintronic devices.

Before analyzing the dynamics of a ferromagnetic macrospin, let us briefly consider a standard electrical LC circuit.
In this description, we will be using the sign convention for voltages and currents as shown in Fig. 1.

We will describe the LC circuit using two dynamical variables: electric charge on the capacitor $Q$ and magnetic flux through the inductor $\Phi$.
These variables are natural variables for the description of the internal state of the capacitor and inductor, respectively.
More importantly, these variables, as it will be demonstrated below, are canonically conjugated variables for the LC circuit, and the equations of motion in these coordinates take the form of standard Hamiltonian equations with the circuit energy $E(Q,\Phi)$ playing the role of the Hamiltonian of the system.

\begin{figure}
\includegraphics[width=2.0in]{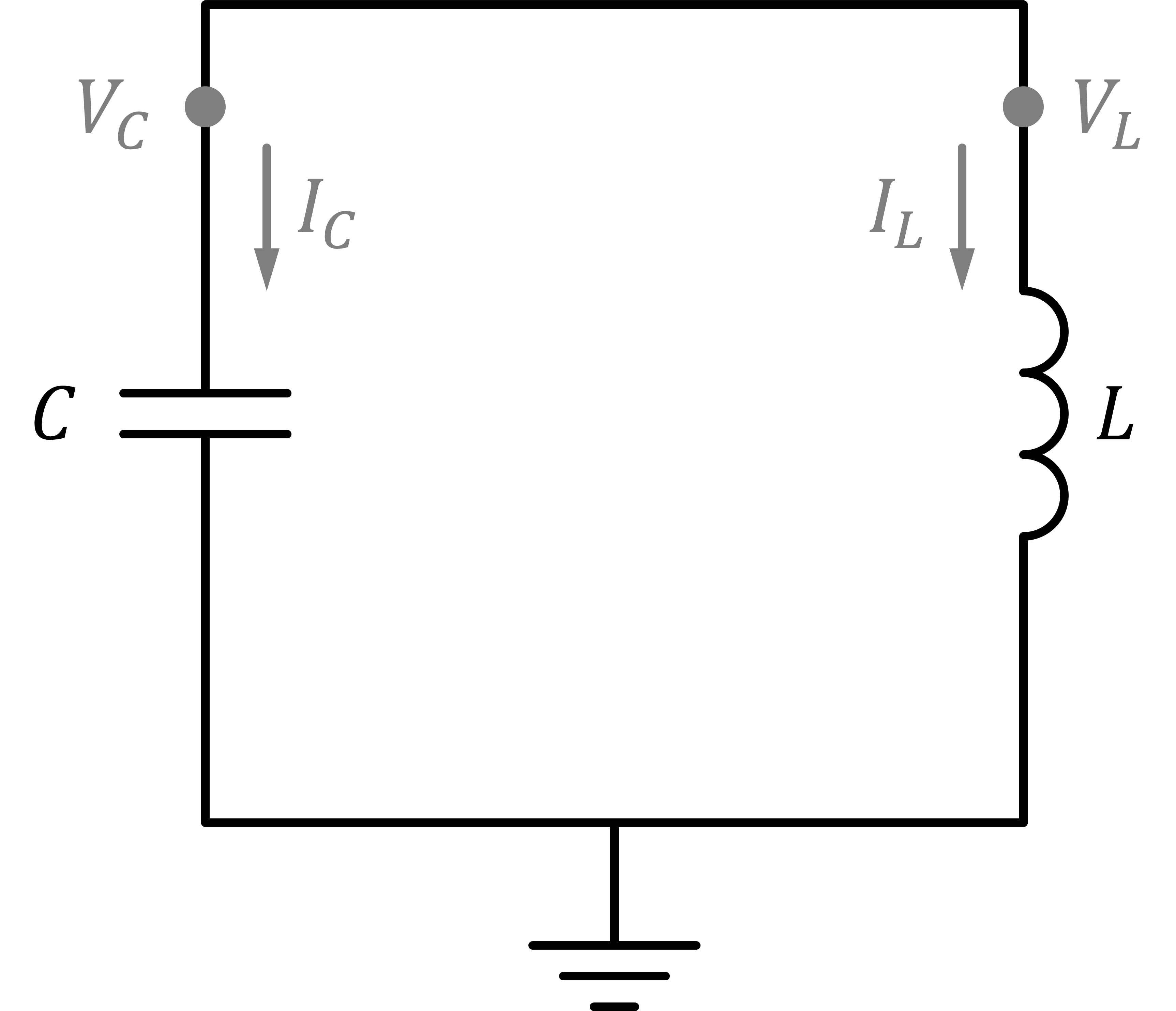}
\caption{\label{rlcircuit} 
Simple LC circuit. 
Nodes indicate capacitance ($V_C$) and inductance ($V_L$) voltages, while arrows show the positive directions of the corresponding currents $I_C$ and $I_L$.
}
\end{figure}

The energy $E(Q,\Phi)$ of the simple linear LC circuit can be written as
\begin{equation}
E(Q,\Phi) = \frac{Q^2}{2C} + \frac{\Phi^2}{2L}.
\label{eq:energy_lc}
\end{equation}
One can use this simple expression to quickly verify the validity of some of the expressions presented below; however, the approach outlined here is valid for an arbitrary nonlinear LC circuit.
In either case, the energy function $E(Q,\Phi)$ completely defines the dynamics of the LC circuit.

Thus, more convenient electrical quantities – current and voltage – are defined through the charge $Q$, flux $\Phi$, and the energy function $E(Q,\Phi)$. 
Namely, the current through the capacitor is, by definition,
\begin{equation}
I_C = \frac{dQ}{dt},
\label{eq:current_cap}
\end{equation}
while the voltage at the capacitor is
\begin{equation}
V_C = \frac{\partial E}{\partial Q}.
\label{eq:voltage_cap}
\end{equation}
Voltage at the inductor is given by Faraday’s induction law,
\begin{equation}
V_L = \frac{d\Phi}{dt},
\label{eq:voltage_ind}
\end{equation}
while the inductor's current is defined as
\begin{equation}
I_L = \frac{\partial E}{\partial \Phi}.
\label{eq:current_ind}
\end{equation}
Combined together, Eqs. \eqref{eq:current_cap}–\eqref{eq:current_ind} leads to the expression
\begin{equation}
\frac{dE}{dt} = V_C I_C + V_L I_L,
\label{eq:energy_rate}
\end{equation}
which is valid for an arbitrary conservative LC circuit, even nonlinear.

For the simple circuit shown in Fig. 1, the dynamical equations follow from Kirchhoff’s current and voltage rules,
\begin{equation}
I_C + I_L = 0,
\label{eq:kirchhoff_current}
\end{equation}
\begin{equation}
V_C - V_L = 0.
\label{eq:kirchhoff_voltage}
\end{equation}
Using Eqs. \eqref{eq:current_cap}–\eqref{eq:current_ind}, these rules can be written as
\begin{equation}
\frac{dQ}{dt} = -\frac{\partial E}{\partial \Phi},
\label{eq:charge_dynamics}
\end{equation}
\begin{equation}
\frac{d\Phi}{dt} = \frac{\partial E}{\partial Q}.
\label{eq:flux_dynamics}
\end{equation}

Equations \eqref{eq:charge_dynamics} and \eqref{eq:flux_dynamics} have the form of standard Hamiltonian equations, with the circuit energy $E(Q,\Phi)$ playing the role of the Hamiltonian of the system. 
This demonstrates that charge $Q$ and flux $\Phi$ are canonically conjugated variables for the LC circuit, with all the advantages provided by classical Hamiltonian dynamics.

For a usual LC circuit. consisting of discrete capacitor and inductor, the energy function has a separable form,
\begin{equation}
E(Q,\Phi) = E_C(Q) + E_L(\Phi),
\label{eq:separable_energy}
\end{equation}
and Eqs.~\eqref{eq:current_cap}–\eqref{eq:current_ind} allow one to derive expressions for the differential capacitance and inductance of the circuit, valid in a general nonlinear case.
Thus, differentiating Eq. \eqref{eq:voltage_cap} with respect to time gives
\begin{equation}
\frac{dV_C}{dt} = \frac{\partial^2 E_C}{\partial Q^2} \frac{dQ}{dt} = \frac{\partial^2 E_C}{\partial Q^2} I_C.
\label{eq:cap_voltage_derivative}
\end{equation}
This equation can be rewritten in the conventional form
\begin{equation}
I_C = C \frac{dV_C}{dt},
\label{eq:cap_current_relation}
\end{equation}
where the differential capacitance $C$ is given by
\begin{equation}
C = \left(\frac{\partial^2 E_C}{\partial Q^2}\right)^{-1}.
\label{eq:capacitance_definition}
\end{equation}
Applying the same analysis to Eq. \eqref{eq:current_ind} gives the conventional relation for the inductor,
\begin{equation}
V_L = L \frac{dI_L}{dt},
\label{eq:ind_voltage_relation}
\end{equation}
with the differential inductance
\begin{equation}
L = \left(\frac{\partial^2 E_L}{\partial \Phi^2}\right)^{-1}.
\label{eq:inductance_definition}
\end{equation}

If the circuit energy does not have a separable form as in Eq. \eqref{eq:separable_energy}, capacitance $C$ and inductance $L$ are not sufficient for a complete description of the circuit.
Proceeding in a similar way, one may derive general relations:
\begin{equation}
\frac{dV_C}{dt} = \frac{\partial^2 E}{\partial Q^2} \frac{dQ}{dt} + \frac{\partial^2 E}{\partial Q \partial \Phi} \frac{d\Phi}{dt} = \frac{\partial^2 E}{\partial Q^2} I_C + \frac{\partial^2 E}{\partial Q \partial \Phi} V_L,
\label{eq:general_voltage_cap}
\end{equation}
and
\begin{equation}
\frac{dI_L}{dt} = \frac{\partial^2 E}{\partial Q \partial \Phi} \frac{dQ}{dt} + \frac{\partial^2 E}{\partial \Phi^2} \frac{d\Phi}{dt} = \frac{\partial^2 E}{\partial Q \partial \Phi} I_C + \frac{\partial^2 E}{\partial \Phi^2} V_L.
\label{eq:general_current_ind}
\end{equation}

These relationships can be rewritten as
\begin{equation}
I_C = C \frac{dV_C}{dt} - \Omega\, C\, V_L,
\label{eq:general_cap_current}
\end{equation}
\begin{equation}
V_L = L \frac{dI_L}{dt} - \Omega\, L\, I_C,
\label{eq:general_ind_voltage}
\end{equation}
where $C$ and $L$ are differential capacitance and inductance, respectively,
\begin{equation}
C = \left(\frac{\partial^2 E}{\partial Q^2}\right)^{-1}, \quad L = \left(\frac{\partial^2 E}{\partial \Phi^2}\right)^{-1},
\label{eq:cap_ind_definitions}
\end{equation}
and $\Omega$ is a new differential circuit parameter defined by
\begin{equation}
\Omega = \frac{\partial^2 E}{\partial Q \partial \Phi}.
\label{eq:omega_definition}
\end{equation}
This parameter has the dimensionality of frequency and describes intrinsic coupling between the charge and flux in the circuit.

In a certain sense, this parameter describes ``memristive'' properties of the circuit, as Eq. \eqref{eq:general_ind_voltage} can be rewritten in the form
\begin{equation}
d\Phi = L \, dI_L + M \, dQ,
\label{eq:memristive_relation}
\end{equation}
reminiscent of the constitutive relation for a memristor, $d\Phi = M \, dQ$.
Here, the ``memristance'' $M = \Omega \, L$ is proportional to the charge-flux coupling parameter $\Omega$.
The memristor analogy, however, is not completely accurate as the charge differential $dQ$ in Eq. \eqref{eq:memristive_relation} is not equal to $-I_L dt$ in a general case.

In this section, we reviewed the charge ($Q$) and flux ($\Phi$) representation of a standard LC circuit. 
This review serves as a foundation for applying similar formalism to macrospin dynamics, where analogous charge and flux variables will be introduced to describe magnetization dynamics.

\section{Charge and Flux Description of a Macrospin}

In this section, we extend the charge-flux representation to a ferromagnetic macrospin, which is a widely used model for describing the magnetization of nanoscale ferromagnets, such as the free layer of a magnetic tunnel junction (MTJ).

We consider a ferromagnetic macrospin (modulus of the gyromagnetic ratio $\gamma$, saturation magnetization $M_s$, and volume $V_o$) precessing under the action of conservative torques (macrospin energy $E = E(\bm{m})$, where $\bm{m}$ is the unit magnetization vector), Gilbert damping (damping constant $\alpha_G$), and spin current $\bm{J}_s(\bm{m})$. 
In a general case, $E(\bm{m})$ and $\bm{J}_s(\bm{m})$ may also depend on time. 
The macrospin dynamics is described by the Landau-Lifshitz-Gilbert-Slonczewski (LLGS) equation, which can be written as
\begin{equation}
\frac{d\bm{m}}{dt} = \gamma \bm{B}_\text{eff} \times \bm{m} + \alpha_G \bm{m} \times \frac{d\bm{m}}{dt} - \frac{\gamma}{ M_s V_o} \bm{m} \times \bm{J}_s \times \bm{m}.
\label{eq:LLGS}
\end{equation}
Here, the first term describes the conservative torque with the effective magnetic field
\begin{equation}
\bm{B}_\text{eff} = -\frac{1}{M_s V_o} \frac{\partial E(\bm{m})}{\partial \bm{m}},
\label{eq:effective_field}
\end{equation}
the second term is the Gilbert damping torque, and the last term is the Slonczewski spin-transfer torque (note that the double cross product with $\bm{m}$ selects the component of $\bm{J}_s(\bm{m})$ that is perpendicular to $\bm{m}$).

Before proceeding further, we shall rewrite Eq. \eqref{eq:LLGS} in a slightly different form. 
Namely, we multiply this equation by the total spin angular momentum of the macrospin,
\begin{equation}
S_s = \frac{M_s V_o}{\gamma},
\label{eq:spin_angular_momentum}
\end{equation}
and take the cross product of both sides with $\bm{m}$. 
This results in the equivalent equation,
\begin{equation}
S_s \bm{m} \times \frac{d\bm{m}}{dt} = -\bm{m} \times \frac{\partial E(\bm{m})}{\partial \bm{m}} \times \bm{m} - \alpha_G S_s \frac{d\bm{m}}{dt} - \bm{m} \times \bm{J}_s (\bm{m}).
\label{eq:modified_LLGS}
\end{equation}
In deriving Eq. \eqref{eq:modified_LLGS}, we used standard vector identities and the orthogonality of $d\bm{m}/dt$ to $\bm{m}$ (which follows from the constant unit length of the vector $\bm{m}$).

There are many ways in which canonically conjugated charge and flux variables can be introduced for a macrospin. 
We consider a version that is convenient for a macrospin precessing close to the $xy$ plane (i.e., an easy-plane macrospin). 
The typical case falling into this category is the free layer of an MTJ that does not have a strong perpendicular magnetic anisotropy (PMA) and is not magnetized to saturation by a perpendicular magnetic field.

Using standard spherical coordinates for the magnetization vector,
\begin{equation}
\bm{m} = \sin\theta \cos\phi \, \bm{e}_x + \sin\theta \sin\phi \, \bm{e}_y + \cos\theta \, \bm{e}_z,
\label{eq:spherical_m}
\end{equation}
the macrospin charge $Q$ and flux $\Phi$ variables are defined as
\begin{equation}
Q = -\frac{e}{\hbar} S_s \cos\theta, \quad \Phi = \frac{\hbar}{e} \phi.
\label{eq:macrospin_charge_flux}
\end{equation}
Here, $e$ is the elementary charge, $\hbar$ is the reduced Planck constant, and the factors like $e/\hbar$ are introduced to convert from ``spin'' to ``electric'' units (i.e., the charge $Q$ has dimensionality of Coulombs, and flux $\Phi$ has dimensionality of Webers). 
As one can see, the charge $Q$ is simply the $z$-component of the spin angular momentum of the macrospin scaled to electrical units. 
One can also see that the charge $Q$ depends only on the polar angle $\theta$, while the flux $\Phi$ depends only on the azimuthal angle $\phi$. 
For this reason, we will often use $\theta$ and $\phi$ as proxies for $Q$ and $\Phi$, respectively.

For the following analysis, it is convenient to introduce two vectors:
\begin{equation}
      \bm{m}_Q  \equiv \frac{\partial{\bm{m}}}{\partial Q}=\frac{\hbar}{e S_s}\left[ \cot\theta (\cos\phi  \bm{e}_x +  \sin\phi  \bm{e}_y) - \bm{e}_z \right],
\end{equation}
\begin{equation}
 \bm{m}_\Phi \equiv \frac{\partial\bm{m}}{\partial\Phi} = \frac{e}{\hbar} \sin\theta \left( -\sin\phi \, \bm{e}_x + \cos\phi \, \bm{e}_y \right).
\label{eq:mPhi}
\end{equation}

Vectors $\bm{m}_Q$ and $\bm{m}_\Phi$ are defined so that they are both orthogonal to each other and to the magnetization vector $\bm{m}$.  
Their norms are given by
\begin{equation}
\bm{m}_Q \cdot \bm{m}_Q = \frac{Z_s}{S_s} \sin^2\theta, \quad \bm{m}_\Phi \cdot \bm{m}_\Phi = \frac{\sin^2\theta}{Z_s S_s},
\label{eq:mQ_mPhi_norms}
\end{equation}
where the constant
\begin{equation}
Z_s = \left(\frac{\hbar}{e}\right)^2 \frac{1}{S_s} = \left(\frac{\hbar}{e}\right)^2\frac{\gamma}{ M_s V_o}
\label{eq:Zs_definition}
\end{equation}
has the dimension of electrical resistance.  
This factor, $Z_s$, naturally appears in the conversion between spin and electrical units and appears in many other expressions below.

Furthermore, one can derive the following cross product relation between $\bm{m}_Q$ and $\bm{m}_\Phi$:
\begin{equation}
\bm{m} \times \bm{m}_Q = \frac{Z_s}{ \sin^2\theta} \bm{m}_\Phi, \quad \bm{m} \times \bm{m}_\Phi = -\frac{\sin^2\theta}{Z_s} \bm{m}_Q.
\label{eq:mQ_mPhi_cross}
\end{equation}
The proportionality factors involve $Z_s$ and $\sin^2\theta$, reflecting the intrinsic coupling between the charge and flux descriptions in the macrospin system.

The most important property that justifies interpreting $Q$ and $\Phi$ as canonically conjugate variables is given by
\begin{equation}
S_s \bm{m}_\Phi \cdot (\bm{m} \times \bm{m}_Q) = 1.
\label{eq:canonical_relation}
\end{equation}
This relation is analogous to the canonical commutation relation found in Hamiltonian mechanics and ensures that $Q$ and $\Phi$ form a proper pair of canonically conjugate variables.  
Such a canonical structure is essential for mapping the macrospin dynamics onto an equivalent electrical circuit model, where charge and flux play roles similar to those of their electrical counterparts.

Now, treating the magnetization $\bm{m}$ as a function of $Q$ and $\Phi$, the LLGS Eq. \eqref{eq:modified_LLGS} takes the form
\begin{equation}
\begin{aligned}
& S_s \bm{m} \times \left[ \bm{m}_Q \frac{dQ}{dt} + \bm{m}_\Phi \frac{d\Phi}{dt} \right]
 = -\bm{m} \times \frac{\partial E}{\partial \bm{m}} \times \bm{m} 
 \\& \quad \quad
 - \alpha_G S_s \left[ \bm{m}_Q \frac{dQ}{dt} + \bm{m}_\Phi \frac{d\Phi}{dt} \right] - \bm{m} \times \bm{J}_s.
\label{eq:LLGS_charge_flux}
\end{aligned}
\end{equation}
Taking the scalar product of this equation with $\bm{m}_\Phi$ and $\bm{m}_Q$, and using the previously derived properties of these vectors, results in the following set of two scalar equations:
\begin{equation}
\frac{dQ}{dt} = -\frac{\partial E(\bm{m})}{\partial \bm{m}} \cdot \bm{m}_\Phi 
-  \frac{\alpha_G\sin^2\theta}{Z_s} \frac{d\Phi}{dt} 
- \bm{m}_\Phi \cdot (\bm{m} \times \bm{J}_s),
\label{eq:Q_dynamics}
\end{equation}
\begin{equation}
\frac{d\Phi}{dt} = \frac{\partial E(\bm{m})}{\partial \bm{m}} \cdot \bm{m}_Q 
+ \frac{\alpha_G Z_s}{ \sin^2\theta} \frac{dQ}{dt} 
+ \bm{m}_Q \cdot (\bm{m} \times \bm{J}_s).
\label{eq:Phi_dynamics}
\end{equation}

Note that the gradient of the macrospin energy with respect to $\bm{m}$ can be expressed in terms of the charge $Q$ and flux $\Phi$ derivatives as follows:
\begin{equation}
\frac{\partial E(\bm{m})}{\partial \bm{m}} \cdot \bm{m}_\Phi = \frac{\partial E(\bm{m})}{\partial \bm{m}} \cdot \frac{\partial \bm{m}}{\partial \Phi} = \frac{\partial E(Q,\Phi)}{\partial \Phi},
\label{eq:energy_derivative_Phi}
\end{equation}
and
\begin{equation}
\frac{\partial E(\bm{m})}{\partial \bm{m}} \cdot \bm{m}_Q = \frac{\partial E(\bm{m})}{\partial \bm{m}} \cdot \frac{\partial \bm{m}}{\partial Q} = \frac{\partial E(Q,\Phi)}{\partial Q}.
\label{eq:energy_derivative_Q}
\end{equation}

The terms with spin current can also be transformed as follows:
\begin{equation}\begin{aligned}
\bm{m}_\Phi \cdot (\bm{m} \times \bm{J}_s) &= -(\bm{m} \times \bm{m}_\Phi) \cdot \bm{J}_s = \frac{\sin^2\theta}{Z_s} \bm{m}_Q \cdot \bm{J}_s \\&= \frac{\sin^2\theta}{Z_s} \frac{\partial P_s}{\partial Q},
\label{eq:spin_current_Q}
\end{aligned}\end{equation}
\begin{equation}\begin{aligned}
\bm{m}_Q \cdot (\bm{m} \times \bm{J}_s) &= -(\bm{m} \times \bm{m}_Q )\cdot \bm{J}_s = -Z_s \bm{m}_\Phi \cdot \bm{J}_s \\&= -\frac{Z_s}{ \sin^2\theta} \frac{\partial P_s}{\partial \Phi},
\label{eq:spin_current_Phi}
\end{aligned}\end{equation}
where the spin current “potential” $P_s$ is defined by
\begin{equation}
P_s = \int \bm{J}_s \cdot d\bm{m}.
\label{eq:spin_potential}
\end{equation}
Note that the integral Eq. \eqref{eq:spin_potential} may not exist for an arbitrary spin current dependence on the magnetization direction. 
This integral, however, exists for a constant $\bm{J}_s$ or for $\bm{J}_s$ having the form $\bm{J}_s(\bm{m}) = J_s (\bm{p} \cdot \bm{m}) \bm{p}, $
where $\bm{p}$ is a constant spin polarization of the spin current. 
These cases correspond to the spin current created by the spin Hall effect and spin currents in driven spin valves and MTJ devices, where $\bm{p}$ is the orientation of the “fixed” magnetic layer.

Using Eqs. \eqref{eq:energy_derivative_Phi}, \eqref{eq:energy_derivative_Q}, \eqref{eq:spin_current_Q}, and \eqref{eq:spin_current_Phi} in combination with Eqs. \eqref{eq:Q_dynamics} and \eqref{eq:Phi_dynamics}, the final form of the equations governing the dynamics of $Q$ and $\Phi$ is obtained:
\begin{equation}
\frac{dQ}{dt} = -\frac{\partial E}{\partial \Phi} -  \frac{\alpha_G\sin^2\theta}{Z_s} \frac{d\Phi}{dt} - \frac{\sin^2\theta}{Z_s} \frac{\partial P_s}{\partial Q},
\label{eq:final_Q_dynamics}
\end{equation}
\begin{equation}
\frac{d\Phi}{dt} = \frac{\partial E}{\partial Q} + \frac{\alpha_G Z_s}{ \sin^2\theta} \frac{dQ}{dt} - \frac{Z_s}{ \sin^2\theta} \frac{\partial P_s}{\partial \Phi}.
\label{eq:final_Phi_dynamics}
\end{equation}
In these equations, both the energy $E$ and the spin current potential function $P_s$ should be treated as functions of $Q$ and $\Phi$.

As one can see, the conservative part of Eqs. \eqref{eq:final_Q_dynamics} and \eqref{eq:final_Phi_dynamics} has exactly the same form as for the standard LC circuit, given by Eqs. \eqref{eq:charge_dynamics} and \eqref{eq:flux_dynamics}. 
This similarity allows us to interpret $Q$ and $\Phi$ as canonically conjugated charge and flux variables in the equivalent electrical model of the macrospin. 
Respectively, the currents and voltages at the equivalent capacitance ($C$) and inductance ($L$) are defined by
\begin{equation}
I_C = \frac{dQ}{dt}, \quad V_C = \frac{\partial E}{\partial Q},
\label{eq:cap_current_voltage}
\end{equation}
\begin{equation}
I_L = \frac{\partial E}{\partial \Phi}, \quad V_L = \frac{d\Phi}{dt}.
\label{eq:ind_current_voltage}
\end{equation}

With these definitions, Eqs. \eqref{eq:final_Q_dynamics} and \eqref{eq:final_Phi_dynamics} can be rewritten as Kirchhoff’s current and voltage rules:
\begin{equation}
I_C + I_L + \frac{V_L}{R_L} = I_s,
\label{eq:kirchhoff_current}
\end{equation}
\begin{equation}
V_C + R_C I_C + V_s = V_L.
\label{eq:kirchhoff_voltage}
\end{equation}

Here, 
\begin{equation}
R_L = \frac{Z_s }{ \alpha_G \sin^2\theta}, \quad R_C = \frac{\alpha_G Z_s}{ \sin^2\theta}
\label{eq:gilbert_resistances}
\end{equation}
are two resistances associated with the Gilbert damping torque, and
\begin{equation}
I_s = -\frac{\sin^2\theta}{Z_s} \frac{\partial P_s}{\partial Q}, \quad V_s = -\frac{Z_s}{ \sin^2\theta} \frac{\partial P_s}{\partial \Phi}
\label{eq:spin_current_sources}
\end{equation}
are current and voltage sources associated with the spin current.

\begin{figure*}[ht]
\includegraphics[width=5.0in]{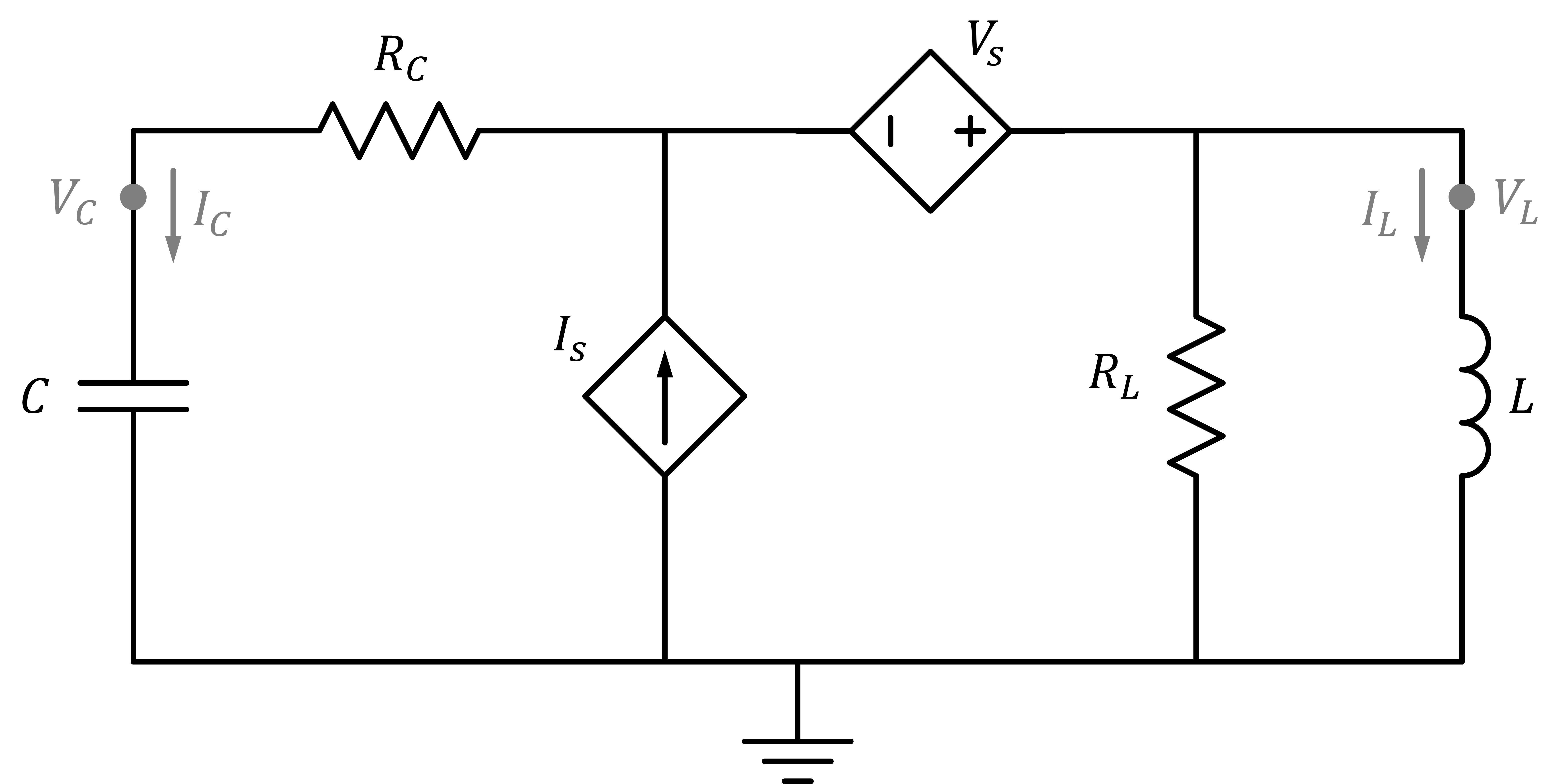}
\caption{\label{bigCircuit} 
Equivalent electric model of a macrospin. 
The capacitor $C$ and inductor $L$ are related to the macrospin energy and conservative torque in the LLGS equation. 
Two resistors, $R_C$ and $R_L$, are associated with the Gilbert damping mechanism. 
The current source $I_s$ and voltage source $V_s$ originate from spin transfer torque. 
All elements of the circuit are, in general, nonlinear and depend on the capacitor charge $Q$ and inductor flux $\Phi$.
}
\end{figure*}

The equivalent circuit described by Eqs. \eqref{eq:kirchhoff_current} and \eqref{eq:kirchhoff_voltage} is shown in Fig.~2. 
The capacitor $C$ and inductor $L$ are related to the macrospin energy $E(Q,\Phi)$ through the constitutive relations given in Eqs. \eqref{eq:cap_current_voltage} and \eqref{eq:ind_current_voltage}. 
The resistors $R_C$ and $R_L$ originate from the Gilbert damping torque and depend on the charge $Q$ through Eqs. \eqref{eq:gilbert_resistances}. 
The spin current-related sources $I_s$ and $V_s$ depend on the particular mechanism of spin current generation and may also depend on the charge $Q$ and flux $\Phi$.

\section{Easy-Plane MTJ}

We now consider a specific case, typical for a free layer of an MTJ without strong perpendicular magnetic anisotropy (PMA). 
Due to a strong demagnetizing field in metallic ferromagnets, such a macrospin can be approximated as an easy-plane ferromagnet (easy plane $xy$, hard axis $z$), where the magnetization almost always remains close to this plane ($\theta = \pi/2$ in spherical coordinates). 
For the following analysis and simulations to remain valid, we require that the magnetization never passes through one of the poles, $\theta = 0$ or $\theta = \pi$. 
We do not impose any restrictions on the range of variation of the azimuthal angle $\phi$, and this model is valid for describing processes such as large-angle out-of-plane precession or magnetization switching.

Formally, we consider the following contributions to the macrospin energy: Zeeman interaction with an arbitrarily oriented external field
\begin{equation}
\bm{B}_e = B_{e,x} \bm{e}_x + B_{e,y} \bm{e}_y + B_{e,z} \bm{e}_z,
\label{eq:external_field}
\end{equation}
hard-axis anisotropy (along the $z$-direction) with anisotropy field $B_d$ (which includes contributions from the demagnetizing and PMA fields), and additional in-plane anisotropy with an easy axis along $x$ and anisotropy field $B_a$. 
Then, the macrospin energy is given by
\begin{equation}
E = M_s V_o \left[ 
- \bm{B}_e \cdot \bm{m} 
+ \frac{1}{2} B_d (\bm{m} \cdot \bm{e}_z)^2 
- \frac{1}{2} B_a (\bm{m} \cdot \bm{e}_x)^2 
\right],
\end{equation}
which can be rewritten using spherical coordinates as
\begin{equation}\begin{aligned}
E =& S_s \gamma \Big[ - (B_{e,x} \cos\phi + B_{e,y} \sin\phi) \sin\theta \\
&- B_{e,z} \cos\theta 
+ \frac{1}{2} B_d \cos^2\theta 
- \frac{1}{2} B_a \cos^2\phi \sin^2\theta 
\Big].
\label{eq:macrospin_energy}
\end{aligned}\end{equation}

In the case where the hard-axis field $B_d$ is much larger than both the bias field $\bm{B}_e$ and the easy-axis field $B_a$, and the magnetization remains close to the equatorial plane $\theta = \pi/2$, the energy function Eq. \eqref{eq:macrospin_energy} can be approximated as
\begin{equation}
E(Q,\Phi) \approx E_C(Q) + E_L(\Phi),
\label{eq:separable_energy_macrospin}
\end{equation}
where
\begin{equation}
E_C(Q) = S_s \gamma \left[ -B_{e,z} \cos\theta + \frac{1}{2} B_d \cos^2\theta \right],
\label{eq:cap_energy_macrospin}
\end{equation}
\begin{equation}
E_L(\Phi) = S_s \gamma \left[ -(B_{e,x} \cos\phi + B_{e,y} \sin\phi) - \frac{1}{2} B_a \cos^2\phi \right].
\label{eq:ind_energy_macrospin}
\end{equation}
In this limit, the charge and flux contributions to the energy become separable, and the macrospin dynamics is equivalent to that of a nonlinear LC circuit with independent capacitance and inductance. 
However, the error introduced by this approximation is of the order of $(|\bm{B}_e| + B_a)/B_d$,
which may be too large for many important cases. 
Therefore, we shall proceed with a more general non-separable form of the energy function given by Eq. \eqref{eq:macrospin_energy}.

Eq. \eqref{eq:macrospin_energy} allows one to find the voltage at the equivalent capacitor,
\begin{equation}\begin{aligned}
V_C = &\frac{\partial E}{\partial Q} = \left(\frac{\hbar}{e}\right) \frac{1}{S_s \sin\theta} \frac{\partial E}{\partial \theta} \\ 
= &\frac{\gamma \hbar}{e} \Big[ -(B_{e,x} \cos\phi + B_{e,y} \sin\phi ) \cot\theta + B_{e,z} \\
&- (B_d + B_a \cos^2\phi ) \cos\theta \Big],
\label{eq:cap_voltage_macrospin}
\end{aligned}\end{equation}
and the current through the equivalent inductor,
\begin{equation}\begin{aligned}
I_L = &\frac{\partial E}{\partial \Phi} = \left(\frac{e}{\hbar}\right) \frac{\partial E}{\partial \phi} \\
= & \frac{\gamma\hbar}{e Z_s} \left[ (B_{e,x} \sin\phi - B_{e,y} \cos\phi) \sin\theta + \frac{1}{2} B_a \sin 2\phi \sin^2\theta \right].
\label{eq:ind_current_macrospin}
\end{aligned}\end{equation}
These equations represent constitutive relations for equivalent nonlinear macrospin capacitance and inductance and can be directly incorporated into modern electrical engineering software (e.g., LTspice) using various behavioral elements.

One can also find the differential characteristics of the equivalent macrospin circuit. 
Thus, the differential capacitance is equal to
\begin{equation}\begin{aligned}
C &= \left(\frac{\partial^2 E}{\partial Q^2}\right)^{-1} \\
&= \left(\frac{1}{\gamma Z_s}\right) \frac{1}{B_d + B_a \cos^2\phi +( B_{e,x} \cos\phi + B_{e,y} \sin\phi)/ \sin^3\theta}.
\label{eq:macrospin_capacitance}
\end{aligned}\end{equation}

The differential inductance is given by
\begin{equation}\begin{aligned}
L &= \left(\frac{\partial^2 E}{\partial \Phi^2}\right)^{-1} \\
&= \left(\frac{Z_s}{\gamma}\right) \frac{1}{(B_{e,x} \cos\phi + B_{e,y} \sin\phi ) \sin\theta + B_a \cos2\phi \sin^2\theta},
\label{eq:macrospin_inductance}
\end{aligned}\end{equation}
and the charge-flux coupling parameter $\Omega$ is
\begin{equation}\begin{aligned}
\Omega &= \frac{\partial^2 E}{\partial Q \partial \Phi} \\
&= \gamma \cos\theta \Big[ B_a \sin2\phi + (B_{e,x} \sin\phi - B_{e,y} \cos\phi ) /\sin\theta \Big].
\label{eq:macrospin_coupling}
\end{aligned}\end{equation}

In the limit of large $B_d$ and small deviations from the easy plane $\left| \theta - \pi/2 \right|$, the equivalent macrospin capacitance becomes constant,
\begin{equation}
C \approx C_0 = \frac{1}{\gamma B_d Z_s},
\label{eq:capacitance_limit}
\end{equation}
the equivalent inductance takes a form similar to the differential inductance of a Josephson junction,
\begin{equation}
L \approx \left(\frac{Z_s}{\gamma} \right)\frac{1}{B_{e,x} \cos\phi + B_{e,y} \sin\phi + B_a \cos2\phi},
\label{eq:inductance_limit}
\end{equation}
while the charge-flux coupling parameter $\Omega$ vanishes, $\Omega \to 0$, indicating the independence of the charge and flux components of the equivalent circuit.


To model the spin current $\bm{J}_s$, we consider a particular case of an unbiased MTJ device with the magnetization of the ``fixed'' layer aligned along the unit vector $\bm{p}$. 
The electrical resistance of such an MTJ can be written as 
\cite{slonczewski1989conductance, wang2009sensitivity}:
\begin{equation}
R_\text{MTJ} = \frac{R_\perp }{ 1 + \eta^2 \cos\beta },
\label{eq:MTJ_resistance}
\end{equation}
where $\eta$ is the dimensionless spin polarization efficiency, and $\beta$ is the angle between the ``fixed'' $\bm{p}$ and ``free'' $\bm{m}$ layer magnetizations, given by
\begin{equation}
\cos\beta = \bm{p} \cdot \bm{m} = (p_x \cos\phi + p_y \sin\phi )\sin\theta + p_z \cos\theta.
\label{eq:MTJ_angle}
\end{equation}
The parameter $R_{\perp}$ represents the MTJ resistance in the perpendicular state ($\beta = 90^\circ$). 
The MTJ parameters $R_{\perp}$ and $\eta$ are related to the MTJ resistance in the parallel ($R_P$) and anti-parallel ($R_{\text{AP}}$) magnetic states by the following relations:
\begin{equation}
R_{\perp} = \frac{2 R_{\text{AP}} R_P}{R_{\text{AP}} + R_P},
\label{eq:MTJ_resistance_perpendicular}
\end{equation}
\begin{equation}
\eta = \sqrt{\frac{R_{\text{AP}} - R_P}{R_{\text{AP}} + R_P}}.
\label{eq:MTJ_polarization}
\end{equation}

The spin current $\bm{J}_s$ for the MTJ can be written in a simple form 
\cite{slonczewski2007theory, sun2008magnetoresistance, wang2009sensitivity}:
\begin{equation}
\bm{J}_s = \frac{\hbar}{e} \frac{\eta}{2} \frac{V_{\text{MTJ}}}{R_{\perp}} \bm{p},
\label{eq:spin_current_MTJ}
\end{equation}
where $V_{\text{MTJ}}$ is the voltage across the MTJ.

Using Eq. \eqref{eq:spin_current_MTJ}, the spin current potential function $P_s(Q,\Phi)$ can be found as
\begin{equation}\begin{aligned}
P_s(Q, \Phi) &= \frac{\hbar}{e} \frac{\eta}{2} \frac{ V_{\text{MTJ}}}{R_{\perp}} \bm{p} \cdot \bm{m} \\
&= \frac{\hbar}{e} \frac{\eta}{2} \frac{V_{\text{MTJ}}}{R_{\perp}} \left[ p_z \cos\theta + (p_x \cos\phi + p_y \sin\phi )\sin\theta \right].
\label{eq:spin_potential_MTJ}
\end{aligned}\end{equation}

Then, the spin current sources $I_s$ and $V_s$ can be written as
\begin{equation}
I_s = \frac{\eta}{2} 
\left[ p_z \sin\theta - (p_x \cos\phi + p_y \sin\phi )\cos\theta \right] \sin\theta \frac{V_{\text{MTJ}}}{R_\perp},
\label{eq:spin_current_source}
\end{equation}
\begin{equation}
V_s = \frac{\eta}{2} \frac{ p_x \sin\phi - p_y \cos\phi }{\sin\theta } Z_s \frac{V_{\text{MTJ}}}{R_\perp}.
\label{eq:spin_voltage_source}
\end{equation}

Equations \eqref{eq:cap_voltage_macrospin}, \eqref{eq:ind_current_macrospin}, \eqref{eq:MTJ_resistance}, and \eqref{eq:spin_current_source}–\eqref{eq:spin_voltage_source} provide a complete description of the equivalent electrical model of an easy-plane MTJ. 
This model can be implemented in electrical engineering software like LTspice using built-in software capabilities.

Note that the MTJ model described here can be easily modified for more specialized cases. 
For example, in the case of an MTJ device with two ``fixed'' layers (e.g., ``polarizer'' and ``analyzer'' layers as in Ref. \cite{houssameddine2007spin}), two spin current sources, each with its own equivalent current $I_s$ and equivalent voltage $V_s$, may be included in the model. 

The case of an MTJ with two ``free'' layers can be modeled using two nonlinear LC circuits (similar to the one shown in Fig. 2) coupled through the spin current sources $I_s$ and $V_s$ and, possibly, through mutual capacitance and inductance, which model the conservative dipolar interaction between the layers.

An example realization of the described here model in LTspice can be found in \cite{MTJSpiceModels} or on the website \href{https://sites.google.com/view/mtjspicemodels/home}{https://sites.google.com/view/mtjspicemodels/home}
while example LTspice-based simulations of several spintronic effects is presented in 
\cite{louis2025physicsbasedcircuitmodelmagnetic}

\vspace{1.2 in}
\null

\section{Conclusion}

In this work, we developed an equivalent electrical model for a ferromagnetic macrospin based on the charge–flux formalism.  
By drawing an analogy to a nonlinear LC circuit, we established a framework in which the macrospin dynamics can be described using capacitance, inductance, and resistive elements, with spin-transfer torque effects incorporated as current and voltage sources.  
This formulation provides a unified perspective that bridges magnetization dynamics and circuit theory, enabling the analysis of spintronic systems using well-established electrical engineering techniques.  
We demonstrated that this approach naturally emerges from the Landau-Lifshitz-Gilbert-Slonczewski (LLGS) equation, reformulating it in terms of charge and flux variables, which allowed us to define an equivalent circuit representation of macrospin dynamics.  

Applying this model to an easy-plane magnetic tunnel junction (MTJ), we showed how the conservative energy contributions and spin-transfer effects can be mapped onto circuit elements.  
The resulting equivalent circuit not only captures the fundamental behavior of the free layer magnetization but also provides a practical means for simulating spintronic devices in software such as LTspice.  
This framework can be further extended to more complex systems, such as MTJs with multiple free layers or devices with additional spin-orbit coupling effects.  
By enabling circuit-based simulations of magnetization dynamics, this work offers a valuable tool for both theoretical studies and practical device design in spintronics.

\bibliography{Der}

\end{document}